\documentclass[11pt]{cernrep}
\usepackage{graphicx}
\usepackage{cite,./mcite}
\usepackage{hyperref}
\usepackage{lineno}
\begin{document}
\title{Data Preservation at LEP}
\author{
Andr\'e G. Holzner$^{1,*}$,
Ryszard Gokieli$^2$, 
Peter Igo-Kemenes$^{1,3}$,
Marcello Maggi$^4$,
Luca Malgeri$^1$, 
Salvatore Mele$^1$, 
Luc Pape$^5$,
David Plane$^1$,
Matthias Schr\"oder$^{1,*}$,
Ulrich Schwickerath$^1$, 
Roberto Tenchini$^6$, 
Jan Timmermans$^7$
}
\institute{
$^1$ CERN, Geneva, Switzerland\\
$^2$ Soltan Institute for Nuclear Studies, Warsaw, Poland\\
$^3$ H{\o}gskolen i Gj{\o}vik, Norway\\
$^4$ INFN Bari, Italy\\
$^5$ ETH Z\"urich, Switzerland\\
$^6$ INFN Pisa, Italy\\
$^7$ NIKHEF, The Netherlands\\
$^*$ corresponding authors}
\maketitle
\begin{abstract}
The four LEP experiments ALEPH, DELPHI, L3 and OPAL successfully recorded
$\mathrm{e^+e^-}$ collision data during the years 1989 to 2000.
As part of the ordinary evolution in High Energy Physics, these experiments can not be repeated
and their data is therefore unique.
This article briefly reviews the data preservation efforts undertaken
by the four experiments beyond the end of data taking. The current status
of the preserved data and associated tools is summarised.
\end{abstract}

\section{Introduction}

This workshop was initiated by the HERA experiments in order
to get a common vision on the issues of data persistency and long term
analysis. In this context, it is interesting to look at how past
experiments dealt with these issues. One example are the LEP
experiments which stopped data taking at the end of the year 2000. 

\section{Preservation of data and software}
The experiments have their data stored in the CASTOR tape archiving
system at CERN. An agreement with the CERN IT division exists so
that these tapes will be copied onto new tapes whenever there is a
migration to new media. 

However, two experiments saw a few tapes being
lost recently\footnote{A suspected source of such tape losses is the fact that LEP data
might coexist with newer and thus more frequently accessed files
(such as from the LHC experiments) on the same tape leading to
unexpected high tape wear.}, so the idea came up to maintain a additional copy of
the LEP data at LHC tier 1 centres. The storage capacity 
needed for this copy (the entire LEP data set is estimated to be 100
TB including Monte Carlo (MC), raw and reconstructed data) would be negligible compared
to the volume of the LHC data and MC productions to come.

Clearly, data is useless without the associated software 
to read and analyse it. The experiments stored their software on AFS at CERN. This includes
reconstruction code as well as simulation code. Analysis codes
are also stored there to various extents. 

The point in time when the simulation and recostruction
codes are considered to be `frozen' is a good occasion to 
collect the components of this part of the overall analysis chain,
package it and make it run on a system independent of the central
software installation. This guarantees to some degree that
no dependencies such as detector conditions databases etc.\ were
forgotten. However, it does not guarantee that this code will
run on recent computing platforms.

DELPHI for example has a software CD project which includes all DELPHI software
in source form. It runs independently from AFS and includes everything
to run the detector simulation and reprocess data, however not the
physics generators.

ALEPH went further and also conserved the computing environment. 
They distributed a `mini-system' to each participating institute
consisting of a laptop and an external disk containing the data and MC
samples. It runs the Linux distribution used at the end of LEP as well 
the necessary ALEPH software to access it. 

OPAL attempts to keep its analysis environment
alive by adapting the software to changes in the computing platform
such as new OS versions and new staging software.
An example of a necessary migration to a different technology is the file catalogue database.
Most of the experiments used FATMEN~\cite{FATMEN} for this purpose. As the central FATMEN server was
decommissioned years ago, the database content had to be moved to
another system, e.g.\ text files (at the cost of some functionality).

All experiments wrote most of their reconstruction and simulation code
in FORTRAN and used ZEBRA~\cite{ZEBRA} (except ALEPH which used BOS~\cite{Blobel:1988gy}) for reading and writing 
event data to files. 
While FORTRAN compilers probably will stay around
for quite some time, the central support for CERNLIB~\cite{CERNLIB} (which ZEBRA is a
part of) has unexpectedly ended and with the upcoming transition 
to version~5 of Scientific Linux CERN, migration and validation
of CERNLIB on the new standard platform will be left to the
experiments. Even though a subset of the CERNLIB functions is now
part of ROOT~\cite{Brun:1997pa}, it is not straightforward to interface them with the existing
experiments' code as this would imply modifications the experiments'
software and the danger of side effects of unknown size.

There are also examples of parts of the software which stopped
working: The event display programs
of OPAL and DELPHI do not run on recent platforms. This is because
they relied on commercial libraries and there is no personpower
available to re-write the complete event display at this stage.

A means against the `ageing' of software is to move to new languages
and libraries along with the operating system and hardware moves.
L3 wrote C++ code which was interfaced with the
existing FORTRAN reconstruction code to read the tracks and
clusters data from ZEBRA files and write them out as ROOT
files. Because ROOT is actively supported today (as opposed to ZEBRA),
this improves the long term accessibility of the data. ALEPH and
DELPHI developed similar C++ frameworks. 


All four experiments investigated QUAERO~\cite{Abazov:2001ny} as a means of
preserving event-by-event data in the form of four-vectors at the level
of reconstructed jets, leptons and photons. The idea was to allow
future physics models beyond the Standard Model to be tested quickly
on preserved data. QUAERO automatically determines the variables built
from four-vector information which are most sensitive to such a model.

Performing such a model-independent analysis turned out to be
difficult as the agreement between MC and data usually was only
checked at some analysis-specific level of preselection. In addition,
tracks and clusters in the event are often interpreted depending
on what is optimal for a given analysis. QUAERO was
therefore not a viable solution for data preservation.


\section{Preservation of knowledge}

While it was straightforward to organise the physical
storage of the data, funding for tapes and their migration being the
only factor, there is no canonical way to ensure 
that the knowledge present in people's minds is written down in any
form of documentation. The range of level of documentation goes
from a complete user's guide to using the collaboration-wide software
to one line descriptions of variables in analysis group specific
ntuples. 

In most cases the last link in the analysis chain, namely
user-specific analysis code and plotting macros, resided in
user's directories and never made it into a centrally managed
repository. 
These directories are deleted by the hosting labs some time after a user has
left the lab, a standard case for highly mobile young students who
often were the responsibles of the analyses, and thus potentially
important information is destroyed.

Information related to the limited precision of the detector 
simulation (e.g.\ which event variables are affected by 
shifts between MC and data and should be used with particular care) 
was probably never systematically collected and written down.
In some cases,
additional smearing of variables in MC was applied in order
to properly describe the data for a given analysis. 
However, such corrections sometimes stayed in the user's analysis code and
were not propagated back into the collaboration's code.

\section{What else to preserve ?}

\begin{figure}
    \centering
    \includegraphics[width=0.9\textwidth]{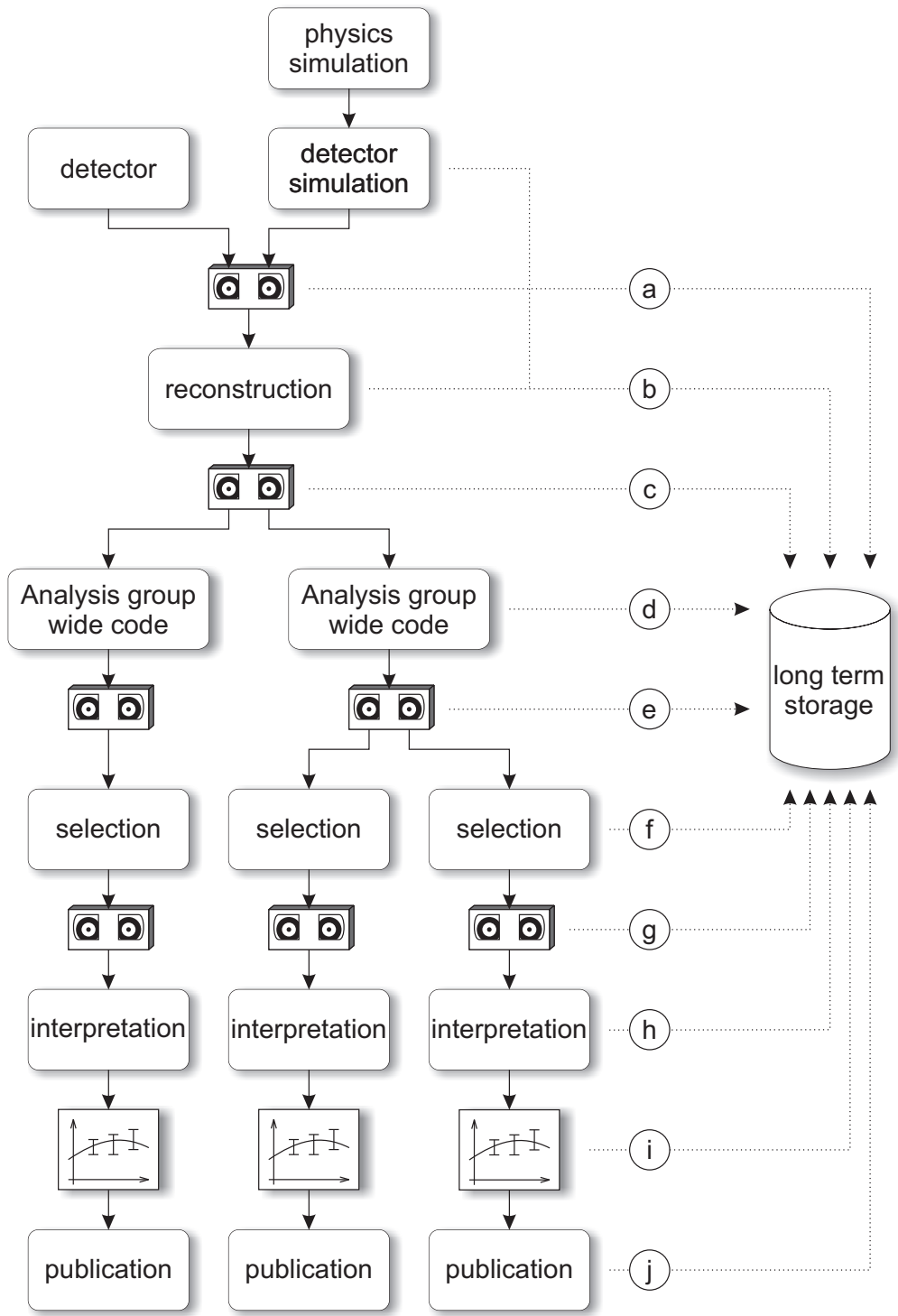}
    \caption{Analysis flow of a typical LEP experiment and possible
      intermediate levels to preserve (dotted arrows)}
    \label{fig:analysis-flow}
  \end{figure}

To get from the reconstructed data to the final publication requires
often the equivalent of several person-years.
Figure~\ref{fig:analysis-flow} shows a typical analysis flow of a
LEP experiment from the recording of the data to the final
publication. For complete preservation, one could think of 
preserving all the intermediate steps (shown in the following table) 
required to get to the publication:

\begin{center}
\begin{tabular}{|c|l|} \hline
Step & Preservation of \\ \hline
a) & raw data\\
b) & detector simulation and reconstruction code\\
c) & reconstruction output (tracks, clusters etc.) \\
d) & tools (`group analysis code') to go from c) to e)\\
e) & analysis groups data\\
f) & tools (`personal analysis code') to go from e) to g)\\
g) & personal n-tuples\\
h) & tools (`macros') to go from g) to i)\\
i) & histograms, systematic uncertainty matrices etc.\\
j) & publications electronically \\
\hline
\end{tabular}
\end{center}

Of course, experiments achieved step a) even though data are threatened
by moving to newer computing platforms. 
In most cases, steps f), g), h) and did not happen. 
Step i) was preserved for some publications,
e.g.~\cite{Achard:2003tx}. A handful of two-dimensional histograms were
published as tables of number of background and data events as well as
signal efficiencies. 
These values have been re-used in a search for new physics~\cite{Achard:2004uu}.
Another example is~\cite{Achard:2004sv}. This 140 page paper contains 66 pages with
77 tables. While with the former paper it is still possible to read
the data from the paper and type it in by hand to use it for
further interpretation, it would be a tedious
and error-prone task to do this for the latter. Fortunately, the
Durham HepData Reaction Database~\cite{durham-hepdata} team
extracted the tables from the electronic
publication
and made them available in a machine-readable format~\cite{Achard:2004sv-hepdata}.
Most experiments made step j) which is implemented by
publishers or {\tt arXiv}.

Going from one abstraction layer of data to the next involves
running computer code. In order to properly use such code one must
rely on accompanying documentation. 
The reconstruction and detector simulation code is usually documented
and thus re-usable because there is a large number of users and most
users are {\it not} developers. This is completely different for
analysis-specific (`personal') code: in most cases the user and the
developer are the same person and thus no effort is spent on documentation.

\section{Collaboration with the other LEP experiments }
All four experiments participated in LEP wide
working groups (WG). The participation usually implied the exchange of some kind of data at
a very high abstraction level 
(but in more detail than shown in the experiments' individual
publications). It also forced the involved persons to write down
documentation on how to use this data.

In some cases, MC files with four-vectors in a common format were produced
(e.g.\ LEP $\mathrm{WW}$ WG~\cite{lep-ww-wg}) or histograms with 
data, signal efficiencies and expected background (LEP SUSY WG~\cite{lep-susy-wg} and LEP
Higgs WG~\cite{lep-higgs-wg})
were exchanged. For the LEP Higgs WG group, some features of the
events observed at highest signal/background were exchanged~\cite{pik-these-proceedings}.


\section{Today's and future use of the data}

ALEPH currently has one ongoing analysis, potentially a few more; 
OPAL has 4-5 papers in the pipeline, L3 has still about five ongoing
analyses while DELPHI has ten papers to be published (for five of them analysis is still ongoing, i.e.\  analysis
jobs running). Thus even years after the collaborations have
disbanded, physics results are still being produced. 

During the active life of the collaboration a paper needed to be approved by the editorial board prior to
publication.
ALEPH has a re-use policy which deviates from this standard procedure~\cite{aleph-policy}. Under this agreement, any
member of the ALEPH collaboration can now publish an analysis
using the archived data without having to go through the full
approval process. Even people outside ALEPH can use ALEPH data for
publications if the author list includes at least one former ALEPH member. Some conditions must be met e.g.\ 
that the analysis does not attempt to reproduce certain
results such as precision measurements. 
About five publications were already published under this scheme
including a re-fit of $\alpha_s$ to
predictions at next-to-next-to-leading order QCD~\cite{Dissertori:2007xa} which only
became available several years after the end of data
taking~\cite{GehrmannDeRidder:2007bj,GehrmannDeRidder:2007hr}. This
analysis could therefore not have been performed during the lifetime of the
collaboration. 
OPAL has a policy similar to the one of ALEPH (i.e.\ outsiders can
join) but still requires approval by the Long-Term Editorial Board~\cite{opal-policy}. 
DELPHI foresees to have a similar policy~\cite{delphi-policy} once the
collaboration is considered to have disbanded.

Should we observe signs of physics beyond the Standard Model in the
future, one wants to test the sensitivty of the LEP data
to such physics models or even confirm or exclude them. The first
step (as suggested in~\cite{opal-policy}) will be to calculate how
such new physics modifies observables measured with great precision at
LEP and compare to the published values. For more in-depth studies,
it will be necessary to run an event generator for the physics
processes in question and very likely this requires the experiments' 
software to run on today's computing platforms.


\section{Conclusions and outlook}

The long term storage of files and associated software was well organised
at the end of data taking. However, due to the shift
from FORTRAN to C++ programming, support for the software libraries 
essential for accessing and analysing data is fading out. 
Thus efforts will be needed to 
assure access to the very valuable LEP data in the future.
Further work is required to preserve (write down) the knowledge
inside the former members' minds.

In case there is a wish or the need to
re-use of LEP data (e.g.\ in light of new theoretical
calculations or signs of new physics at running or future
experiments), there are still some collaboration members around to do this.
ALEPH and OPAL (and most likely DELPHI in the future as well) even allow the use of their
data by `outsiders' under the guidance of a former collaboration member.
%

\bibliographystyle{dplta} 
{\raggedright
\bibliography{dplta}
}
\end{document}